\UseRawInputEncoding 
\documentclass[aps,prl,twocolumn,superscriptaddress,floatfix]{revtex4-2}

\usepackage[
  separate-uncertainty = true,
  multi-part-units=single
]{siunitx}

\usepackage{amsmath,amssymb}
\usepackage{graphicx}
\usepackage{color}
\usepackage{xcolor}
\usepackage{ulem}
\usepackage{bm}
\usepackage[colorlinks=true,linkcolor=blue,allcolors=blue]{hyperref}
\usepackage{float}

\usepackage{subfigure}

\begin{document}


\title{Direct Measurement of Curvature-Dependent Surface Tension in a Capillary-Condensed Alcohol Nanomeniscus}


\author{Dohyun Kim}
\affiliation{Center for 0D Nanofluidics, Institute of Applied Physics, Department of Physics and Astronomy, Seoul National University, Seoul 08826, Republic of Korea.}
\author{Jongwoo Kim}
\affiliation{Center for Convergent Research of Emerging Virus Infection, Korea Research Institute of Chemical Technology, Daejeon 34114, Republic of Korea.}
\author{Jonggeun Hwang}
\affiliation{Center for 0D Nanofluidics, Institute of Applied Physics, Department of Physics and Astronomy, Seoul National University, Seoul 08826, Republic of Korea.}
\author{Dongha Shin}
\affiliation{Center for 0D Nanofluidics, Institute of Applied Physics, Department of Physics and Astronomy, Seoul National University, Seoul 08826, Republic of Korea.}
\affiliation{Division of Fine Chemistry and Engineering, College of Natural Science, Pai Chai University, Daejeon 35345, Republic of Korea.}
\author{Sangmin An}
\affiliation{Center for 0D Nanofluidics, Institute of Applied Physics, Department of Physics and Astronomy, Seoul National University, Seoul 08826, Republic of Korea.}
\affiliation{Department of Physics, Research Institute of Physics and Chemistry, Jeonbuk National University, Jeonju 54896, Korea.
}
\author{Wonho Jhe}
\email[corresponding author.]{whjhe@snu.ac.kr}
\affiliation{Center for 0D Nanofluidics, Institute of Applied Physics, Department of Physics and Astronomy, Seoul National University, Seoul 08826, Republic of Korea.}


\begin{abstract}
Surface tension is a key parameter for understanding nucleation from the very initial stage of phase transformation. Although surface tension has been predicted to vary with the curvature of the liquid-vapor interface, particularly at the large curvature of, e.g., the subnanometric critical nucleus, experimental study still remains challenging due to inaccessibility to such a small cluster. Here, by directly measuring the critical size of a single capillary-condensed nanomeniscus using atomic force microscopy, we address the curvature dependence of surface tension of alcohols and observe the surface tension is doubled for ethanol and \textit{n}-propanol with the radius-of-curvature of $\sim$\SI{- 0.46}{nm}. We also find that the interface of larger negative (positive) curvature exhibits the larger (smaller) surface tension, which evidently governs nucleation at $\sim$\SI{1}{nm} scale, indicating more facilitated nucleation than normally expected. Such well characterized curvature effects contribute to better understanding and accurate analysis of nucleation occurring in various fields including material science and atmospheric science.
\end{abstract}

\keywords{Surface tension, Alcohol, Nanomeniscus, Heterogeneous nucleation, Curvature dependent surface tension, Tolman length}

\date{\today}

\maketitle


Nucleation of alcohols is a fundamentally important process since it plays a critical role in investigating the atmospheric new particle formation, which affects earth radiation budget, climate change and human health \cite{Stolzenburg2018,Kulmala2013,Andreae2008,Shiraiwa2017}. 
To study formation and growth of aerosol particles from the initial stage of nucleation, alcohols are widely employed to activate and help grow the fine molecular particles to detectable sizes, such as used in condensation particle counters \cite{Kulmala2012,Dada2020}. 
In addition, alcohols are a simple type of organic compounds that serve as major contributors to new particle formation in atmosphere \cite{Stolzenburg2018}.
Therefore, quantitative understanding of nucleation of alcohols is essential from laboratory experiments to field studies that address the mechanism of atmospheric new particle formation as well as the activation properties of those particles and vapors, which are attracting increasingly more attention as concerns on the environmental issues such as air pollution increase \cite{Winkler2008,Yao2018,Guo2020}.

\begin{figure}[ht]
\subfigure{\label{subfig-1}}\subfigure{\label{subfig-2}}\subfigure{\label{subfig-3}}\subfigure{\label{subfig-4}}\subfigure{\label{subfig-5}}\subfigure{\label{subfig-6}}
\includegraphics[width=8.5cm]{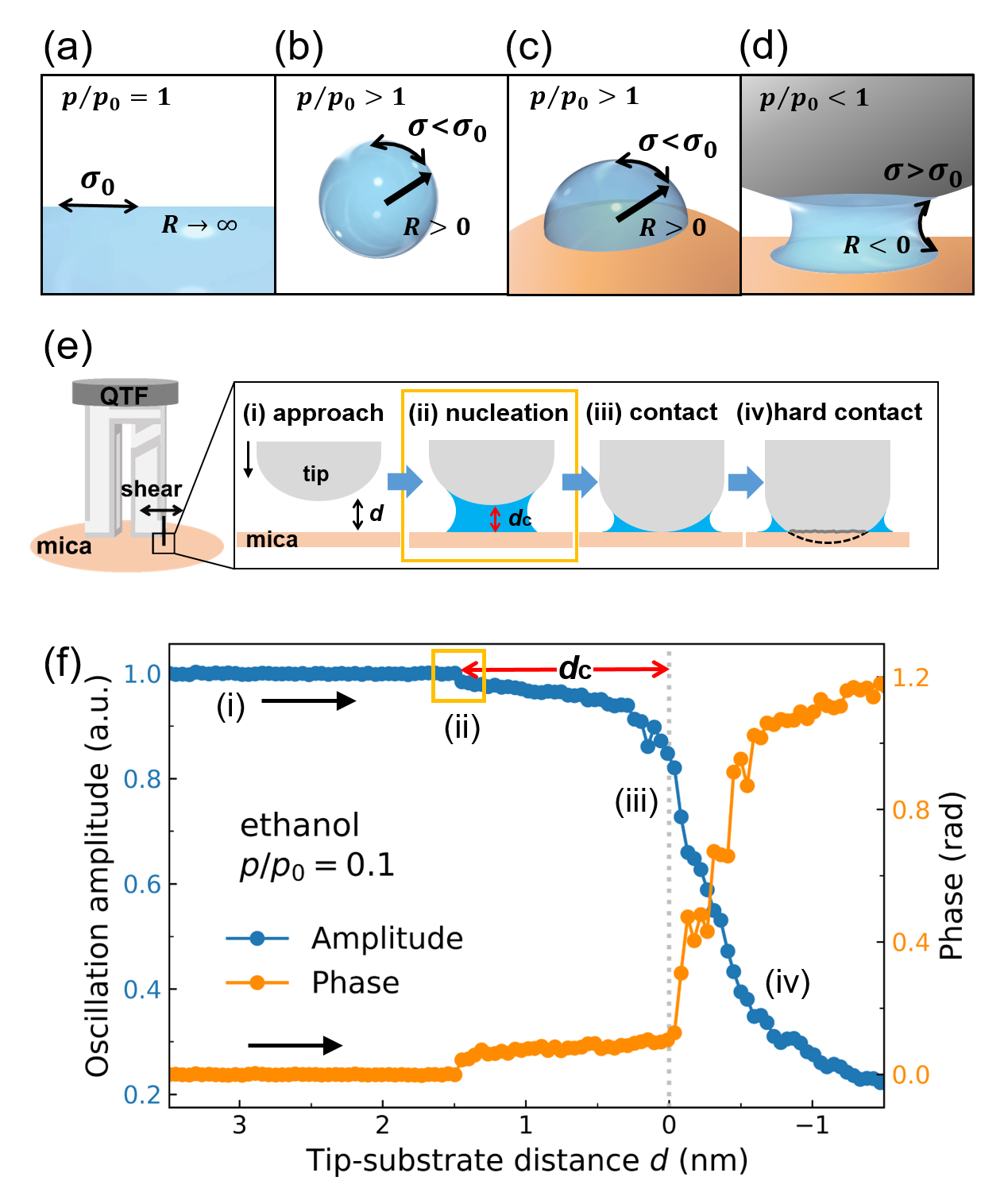}
\caption{\label{fig:Fig1} (a) In saturated vapor ($p_{}/p_{0}$ = 1), the liquid-vapor (LV) interface is planar in equilibrium with constant surface tension $\sigma_{0}$. (b), (c) In supersaturated vapor, the LV interface has a positive curvature in equilibrium, irrespective of a solid interface. (d) In unsaturated vapor, the LV interface of a meniscus has a negative curvature. Notice that surface tension $\sigma$ varies depending on the curvature in (b)$\sim$(d) unlike in (a). (e) A fused quartz tip is glued on a QTF prong for shear-mode AFM. (f) Amplitude and phase signals of QTF obtained as the tip approaches mica. Capillary condensation occurs at the critical distance $d_\mathrm{c}$. (i)-(iv) show each step of tip approach, allowing direct measurement of $d_\mathrm{c}$ and contact point ($d$ = 0).}
\end{figure}

When characterizing phase transitions including such nucleation process, surface tension is an important thermodynamic parameter \cite{Ovadnevaite2017,ALBERTI2019,Kelton2010}. It determines the energy barrier to overcome to form a cluster of new phase and also provides the equilibrium criteria for the critical nucleus, the very first stage of nucleation. However, surface tension is expected to vary depending on the curvature of the interface (Fig.~\ref{subfig-1} to \ref{subfig-4}), particularly for a very small cluster down to sub-nanometer scale \cite{Tolman1948,Tolman1949}. Despite numerous simulation and theoretical studies, the curvature-dependent surface tension is still an issue under ongoing debate \cite{Anisimov2007,Lei2005,Rehner2019}. Especially for alcohols, there have been studies that incorporate the curvature correction to the alcohols' surface tension for better analysis of the results in the homogeneous nucleation (HON) rate experiment \cite{Bruot2016,Asen2020}. Nevertheless, there is no direct measurement of surface tension of alcohols for sufficiently large curvature of interface, or extremely small size, to clearly identify its curvature dependency.
Because the nucleation-rate measurement for alcohols has been conducted in supersaturated alcohol vapor, it produces the nuclei of positive-curvature interface in equilibrium (Fig.~\ref{subfig-2} and \ref{subfig-3}), which are unstable and thus lead to spontaneous growth immediately after nucleation
\cite{Strey1986,Grabmann2002,Zhang2010}. In such systems, it is impossible to observe directly a nanoscale critical nucleus, which can be detected only when grown enough. Therefore, for better understanding of nucleation of alcohol, the main subject of the present work, one needs to investigate the surface tension by direct size measurement of the critical nucleus in non-supersaturated condition (Fig.~\ref{subfig-4}).

Here, we quantify the curvature dependency of surface tension of ethanol and \textit{n}-propanol at the nanoscale by direct size measurement of a single critical nucleus capillary-condensed between two surfaces. Since the experiment is conducted at unsaturated low vapor pressure ($p_{}/p_{0}$ $<$ 0.5), a nanomeniscus with large negative curvature is formed (Fig.~\ref{subfig-4}), which is (meta-)stable in equilibrium and thus minimizes the typical systematic errors in size characterization, in contrary to the supersaturated case \cite{Restagno2000,Men2009,Zhang2014}. Note that the curvature dependency for nucleation systems having either signs of curvature (Fig.~\ref{subfig-2} to~\ref{subfig-4}) is closely interrelated with each other \cite{Melrose1968}.

For simultaneous formation and direct observation of a critical nucleus having large negative curvature, we approach a sharp quartz tip to mica substrate in ambient condition, employing quartz tuning fork-based atomic force microscopy (QTF-AFM). The AFM operates in shear mode with the tip oscillating laterally, allowing precise measurement of the tip-substrate distance (Fig.~\ref{subfig-5}).
During tip approach, a nanomeniscus-shaped nucleus is produced in the gap at the critical tip-surface distance of nucleation, $d_\mathrm{c}$, which characterizes the size of a critical nucleus and relates uniquely to its equilibrium curvature (inset of Fig.~\ref{subfig-5}). The oscillation amplitude and phase responses of QTF are monitored during approach (Fig.~\ref{subfig-6}).
Since QTF is a very sensitive and stiff oscillator, QTF-AFM is suitable to detect $in$ $situ$ the single critical nucleus (Fig.~\ref{subfig-6} (ii)) and to control the tip-surface distance accurately without mechanical instability. Details of measurement are reported in previous works \cite{Choe2005,Lee2007,Kim2018} and Supplemental Material (SM) 1.

The liquid-vapor (LV) interface of the formed nanomeniscus can be described by the Kelvin equation, which relates the equilibrium curvature with the vapor pressure,

\begin{equation}
\frac{1}{R}=\frac{k_B T}{v_{l} \, \sigma} \ln{\frac{p}{p_0}},
\label{eq:one}
\end{equation}

where $R$ is the mean radius-of-curvature of the interface, $k_B$ the Boltzmann constant, $T$ temperature, $v_l$ molecular volume of liquid, $\sigma$ surface tension, and $p_{}/p_{0}$ ratio of external vapor pressure to saturation pressure. The Kelvin equation shows that for $p_{}/p_{0}$ $<$ 1, lower $p_{}/p_{0}$ produces interface with larger absolute value of curvature $1/|R|$ in equilibrium, whereas for $p_{}/p_{0}$ $>$ 1, larger $1/R$ is produced at higher $p_{}/p_{0}$. 
Notice that Eq.~\ref{eq:one} assumes constant surface tension and its validity at the nanometric scale is still in debate \cite{Kohonen2000,Factorovich2014}. It is important to remark that the equilibrium mean curvature ($1/R$) of the nanomeniscus obtained at certain $p_{}/p_{0}$ is uniquely determined by $d_\mathrm{c}$, above which no solutions exist for the Young-Laplace equation with given tip radius and contact angles \cite{Orr1975} (See SM 2). This justifies the direct and accurate measurement of $d_\mathrm{c}$ where nucleation occurs, providing the unique platform to make measurements on the curvature-dependent surface tension.

\begin{figure*}
\subfigure{\label{subfig-7}}\subfigure{\label{subfig-8}}\subfigure{\label{subfig-9}}\subfigure{\label{subfig-10}}\includegraphics[width=17.5cm]{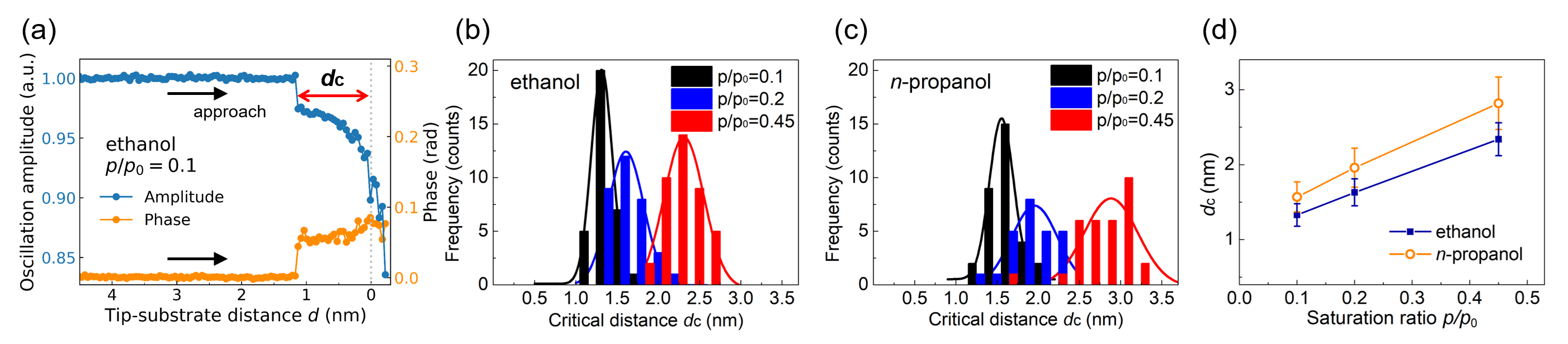}\caption{\label{fig:Fig2} (a) Typical amplitude and phase for ethanol at $p_{}/p_{0}$ = 0.1. The tip approaches in steps of \SI{0.05}{nm} and stays for 500 ms at each step. (b) Histogram plot of $d_\mathrm{c}$ versus $p_{}/p_{0}$ obtained for 30 to 39 times tip approach in ethanol vapor and (c) for 25 to 33 times measurement in \textit{n}-propanol vapor. (d) Larger $d_\mathrm{c}$ at larger $p_{}/p_{0}$. $d_\mathrm{c}$ is measured as \SI{1.33 \pm 0.15}{nm}, \SI{1.63 \pm 0.18}{nm} and \SI{2.34 \pm 0.22}{nm} at $p_{}/p_{0}$ = 0.1, 0.2 and 0.45, respectively, for ethanol and \SI{1.57 \pm 0.2}{nm}, \SI{1.96 \pm 0.26}{nm} and \SI{2.82 \pm 0.35}{nm} at each $p_{}/p_{0}$ for \textit{n}-propanol. Each error denotes the standard deviation and the solid lines are for eye guide.}
\end{figure*}

The typical approach curves are presented in Fig.~\ref{subfig-7} for ethanol (See Fig.~S2 for \textit{n}-propanol).
The histograms and Gaussian fittings for measured values of $d_\mathrm{c}$ are shown in Figs.~\ref{subfig-8} and \ref{subfig-9}, obtained by 25 to 39 independent measurements at three different saturation ratios for each alcohol. As summarized in Fig.~\ref{subfig-10}, larger $d_\mathrm{c}$ is obtained at larger $p/p_0$ for both ethanol and \textit{n}-propanol, for which the general tendency follows Eq.~\ref{eq:one}.
Notice that for accurate measurements of $d_\mathrm{c}$, we carefully deal with the contamination issue and the liquid film leftover on the surfaces: To prevent contamination of tip and substrate, the tip is retracted before reaching hard contact (Fig.~\ref{subfig-5}(iv)) as in Fig.~\ref{subfig-7}, protecting tip from deformation. To avoid overestimation of $d_\mathrm{c}$ due to the leftover liquid films on mica surface \cite{Bampoulis2016,Loi2002,Gee1989}, each measurement is conducted at different locations of about $\SI{100}{nm}$ apart on mica. Detailed analysis of the factors that can affect the measurement is in SM 3.

\begin{figure}[ht]
\subfigure{\label{subfig-13}}\subfigure{\label{subfig-14}}\includegraphics[width=8.5cm]{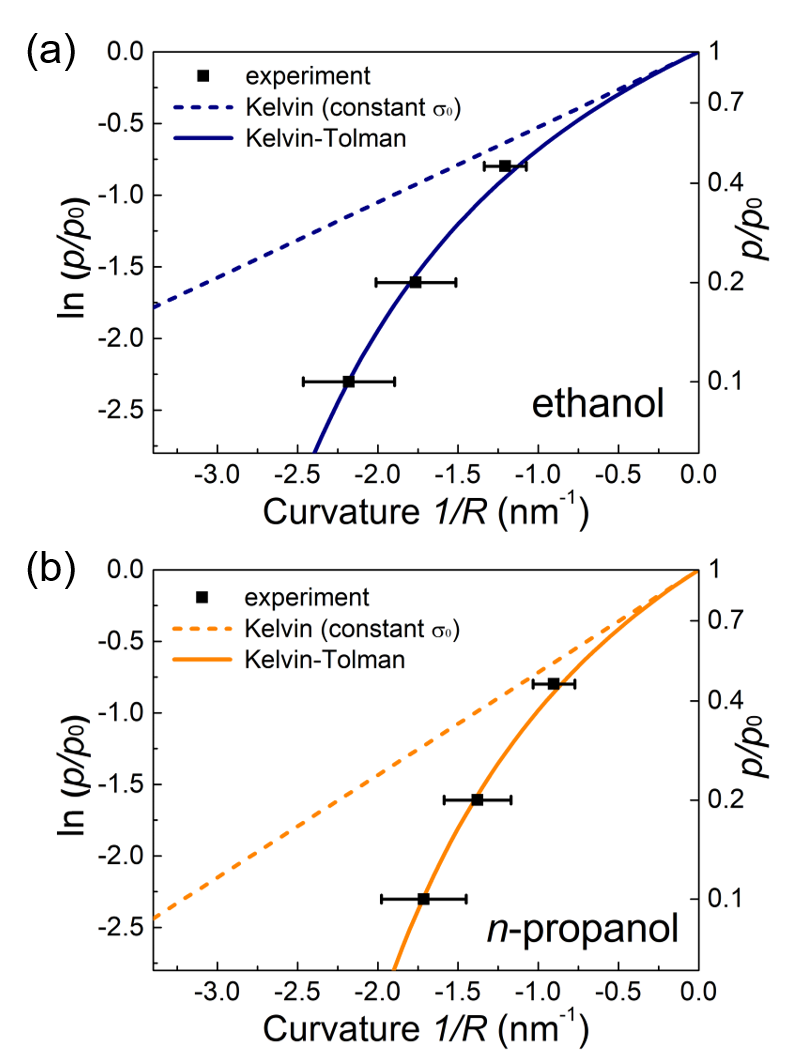}
\centering
\caption{\label{fig:Fig3} Plot of $p_{}/p_{0}$ versus the equilibrium curvature $1/R$. Black squares with error bars represent the $1/R$ values obtained from the measured $d_\mathrm{c}$. The solid curves represent the fits of the first-order Kelvin-Tolman equation on the experimental data, showing (a) $\delta=\SI{0.23}{nm}$ for ethanol and (b) $\delta=\SI{0.27}{nm}$ for \textit{n}-propanol, respectively, whereas the dashed lines indicate the Kelvin equation that assumes constant surface tension.}
\end{figure}

Figure~\ref{fig:Fig3} presents the results of the exact $1/R$ versus $p_{}/p_{0}$, where the plotted values of $1/R$ (black squares) correspond uniquely to each measured $d_\mathrm{c}$. As observed clearly, experimental data show considerable discrepancies with the simple $1/R$ behavior predicted by Eq.~\ref{eq:one} (dashed lines), where we use the well-known constant surface tension for ethanol ($\SI{22.01}{mN/m}$) and \textit{n}-propanol ($\SI{23.5}{mN/m}$) in Figs.~\ref{subfig-13} and \ref{subfig-14}, respectively. As shown, larger discrepancies between the observed and predicted values occur at lower saturation ratio $p_{}/p_{0}$. It implies that the curvature-dependent surface tension has to be incorporated, since careful consideration of other factors, including nonideality of the fluid and contact-angle change due to the line-tension effect at the three-phase contact line \cite{Hienola2007,Winkler2016,Masao2019}, fails to fully explain the discrepancies. Especially, the discrepancy actually increases with contact angle increase within the range $0<\theta<\pi/2$, but decreases for vanishing contact angle. Even when zero contact angles are assumed both between tip and meniscus and between meniscus and mica, considerable discrepancies still prevail. See SM 4 for details. 

\begin{figure}[ht]
\includegraphics[width=8.5cm]{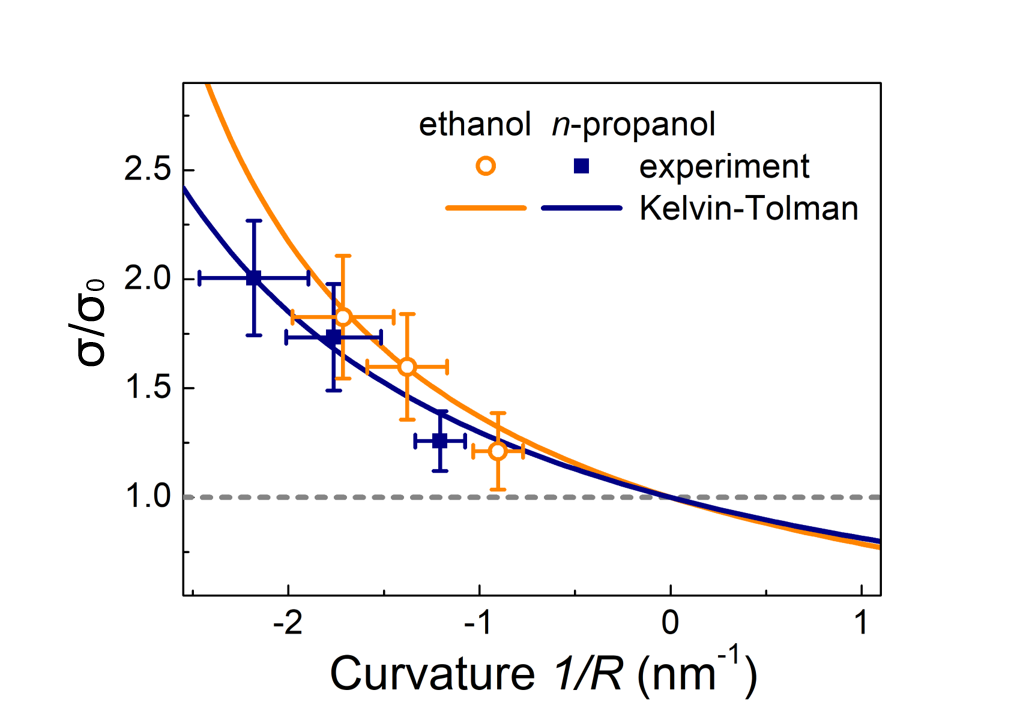}
\centering
\caption{\label{fig:Fig4} The normalized ratio of surface tension $\sigma/\sigma_0$. The larger surface tension is obtained for smaller nanomeniscus, which equivalently implies that the smaller nanodroplet of positive curvature has smaller surface tension.}
\end{figure}

The curvature-dependent surface tension has been characterized by the Tolman length $\delta$, as given by the Tolman equation \cite{Tolman1949},

\begin{equation}
\sigma_{}=\frac{\sigma_0}{1+\frac{\delta}{R}},
\label{eq:two}
\end{equation}

where $\sigma_{0}$ is the bulk surface tension at planar interface. Though it was first derived for a spherical droplet, the Tolman equation is extended to a general shape of interface beyond a sphere \cite{Melrose1968}. Combining Eqs.~(\ref{eq:one}) and (\ref{eq:two}) leads to the following Kelvin-Tolman equation,

\begin{equation}
\frac{1}{R}=\frac{k_B T}{v_{l} \, \sigma_0} \left(1+\frac{\delta_{}}{R}\right) \ln{\frac{p}{p_0}},
\label{eq:three}
\end{equation}

which is the first-order approximation of curvature dependency assuming constant $\delta$. 
The solid curves in Fig.~\ref{fig:Fig3} represent the fits of Eq.~\ref{eq:three} to the experimental data using $\delta$ as a fitting parameter. We find $\delta$ is about \SI{0.23}{nm} for ethanol and \SI{0.27}{nm} for \textit{n}-propanol, both with positive sign, indicating the larger surface tension for smaller meniscus. Fig.~\ref{fig:Fig4} shows such surface-tension variations for both alcohols. For example, for ethanol at $R\approx\SI{-0.46}{nm}$, surface tension is twice the bulk value.

Notice that the errors of $1/R$ in Figs.~\ref{fig:Fig3} and \ref{fig:Fig4} correspond to the error of $\delta$ up to $\SI{\pm 0.09}{nm}$ for ethanol and $\SI{\pm 0.16}{nm}$ for \textit{n}-propanol.
Although each error range of $\delta$ somewhat overlaps (Fig.~\ref{fig:Fig4}), the fact that $\delta_\mathrm{ethanol}$ < $\delta_\mathrm{propanol}$ reflects the slight difference in their molecular size and shape. Because, based on its definition, the value of $\delta$ is known to be closely related to the intermolecular distance and interaction range \cite{Lei2005}.
In addition, quantitative analysis of possible errors for other quantities used in the calculation shows that even when $\delta$ varies by $-33 \%$ up to $+78 \%$, the sign of $\delta$ is invariant (See SM 4).

Interestingly, the positive sign of $\delta$ indicates that a cluster of condensed phase with more positive (negative) curvature $1/R$ has smaller (larger) surface tension $\sigma$ than the bulk value $\sigma_0$ at the planar interface (solid curves in Fig.~\ref{fig:Fig4}). This suggests that alcohols can nucleate more frequently, or equivalently, activate at a lower vapor pressure than what the classical nucleation theory (CNT) predicts. This is because although CNT has been considered as a qualitatively correct way to describe the critical nucleus and to predict the nucleation rate, it is based on Eq.~\ref{eq:one}, assuming constant surface tension. Our results are also consistent with the previous studies on the positive curvature case, including the HON rate measurement and nucleating colloidal liquid, all reporting the lower surface tension than bulk to properly explain their data \cite{Bruot2016,Asen2020,Nguyen2018}. Notice that an even better agreement appears when the second-order correction is implemented by using the curvature dependence of the Tolman length itself (SM 5).

Moreover, it can be employed to the heterogeneous nucleation (HEN) case because Eq.~\ref{eq:one} is established for LV interface regardless of the presence of solid surface \cite{Yarom2015}. The previous studies on the HEN rate measurement have shown the same tendency, as ours, that the onset saturation ratio $p/p_0$ is lower than what the CNT with constant surface tension expects \cite{Hienola2007,Winkler2008}. Since these experiments that form convex clusters of \textit{n}-propanol on pre-existing particles (Fig.~\ref{subfig-3}) mainly focus on the size of the activated particles and the liquid-solid interface properties rather than the LV interface, surface tension change has not been seriously considered even for sufficiently large curvature. In fact, the slight discrepancy that still exists between HEN experiment and the theory \cite{Winkler2008} can be quantitatively accounted for by incorporating the curvature-dependent surface tension, as discussed in SM 6. Thus, we have demonstrated that our results on curvature-dependent surface tension may be helpful in further studies on nucleation and subsequent growth of molecular-scale particles in various fields such as material science, atmospheric science, physics and chemistry. Moreover, surface tension of alcohols may serve as a practically important parameter for accurate analysis and interpretation in those studies since only $\sim 20\%$ change in surface tension leads to orders-of-magnitude change in the nucleation rate prediction \cite{Bruot2016,Nguyen2018,Ovadnevaite2017}. 

In summary, the curvature-dependent surface tension of ethanol and \textit{n}-propanol, showing up to twofold enhancement, is obtained by directly measuring the size of critical nucleus with curvature up to $\sim \SI{-2.2}{nm^{-1}}$ and $\SI{-1.7}{nm^{-1}}$, respectively. In addition, the Tolman length, a parameter characterizing the curvature dependency, for ethanol is found to be $\sim 20\%$ smaller than that for $n$-propanol, reflecting the relevant difference of molecular structure.
Exact knowledge of the curvature dependency of pure fluid's surface tension provides a crucial basis for analyzing the binary, ternary, and multi-component cases, which is necessary for understanding and characterizing the nucleation processes in complex real systems from atmospheric to biological systems \cite{Kulmala2013,Stolzenburg2018,ALBERTI2019}.


%


\begin{acknowledgements}

The SEM images of the quartz tip were obtained at the Research Institute of Advanced Materials, Seoul National University. The authors acknowledge Jongsook Kim for operating SEM. This work was supported by the National Research Foundation of Korea (NRF) grant funded by the Korea government (MSIP) (No. 2016R1A3B1908660), by Basic Science Research Program through the National Research Foundation of Korea (NRF) funded by the Ministry of Education, Science and Technology (2020R1I1A1A01070755), and by a National Research Council of Science and Technology (NST) grant from the Korean Government (MSIT) (CRC-16-01-KRICT).

\end{acknowledgements}

\bibliography{reference.bib}

\end{document}